\documentclass{ws-procs9x6}

\begin{document}

\title{non-linear waves in fluids\\ near the critical point}

\author{H. GOUIN}

\address{{Laboratoire de Mod{\'e}lisation en M{\'e}canique et
Thermodynamique,} \\
E.A. 2596, Universit{\'e} d'Aix-Marseille, 13397 Marseille Cedex 20, France \\
E-mail: henri.gouin@univ.u-3mrs.fr}

\maketitle

\abstracts{A non linear model associated with a Landau-Ginzburg-like
behavior in mean field approximation forecasts phase transition
waves and solitary kinks near the critical point. The behavior of
isothermal waves is  different of the one of isentropic waves as
well in conservative cases as in dissipative cases. }

\section{Introduction}

The study of plane waves propagating in   fluid second gradient
theory is a main subject in continuum mechanics
\cite{Germain}$^-$\cite{Gouin05}. From our paper
 (\cite{Gouin93})  where
 strong variation of density for the matter
constituent occurs in isothermal conservative motions, we propose an
extension to other thermodynamical conditions and for dissipative
motions. Such cases appear in phase transitions through interfacial
layers \cite{Widom,Rowlinson}. Near the critical point the thickness
of the interface separating the two phases of
 a fluid gets non molecular dimension. The theory of second gradient
corresponding to an extended van der Waals model with
thermodynamical potentials modeling  the mean field approximation
allows to study the interface in equilibrium
\cite{vanderWaals,Cahn}. With a convenient rescalling, the governing
equations of material waves can be expressed   for conservative and
viscous motions with a Landau-Ginzburg-like behaviour \cite{Landau};
they are compatible with the second law of thermodynamics and we can
rewrite the one-dimensional equations of non homogeneous fluids in
terms of density \cite{Casal 1985a,Casal1988}. To obtain wave
motions, we study the thermodynamical potentials of a fluid near the
critical point and extend the expression given in
\cite{Widom,Rowlinson}
 to the local internal energy. Thermodynamical potentials are
 introduced in equation of motions in two cases:
  when the temperature is fixed
(isothermal motions) and when the specific entropy along
trajectories is constant (isentropic motions). We find the
possibility of non-linear waves in form of solitary waves or phase
transition waves for conservative motions, but no kinks as traveling
waves can appear for non conservative  motions. Finally we discuss
the fluid velocity near the critical point depending on the boundary
conditions.

\section{Thermodynamic potentials near the critical point}
In mean field approximation, the thermodynamical potentials of
 fluids can be expressed in an analytical form near the
critical point \cite{Rowlinson}. The chemical potential of a
homogeneous fluid can be written in the following form
\cite{Widom,Swinton}
\begin{equation}
\mu = \mu^c + \mu^c_{01}\, (T-T_c) + \mu^c_{11}(\rho-\rho_c)\ (T-T_c)
+ \frac{1}{6}\  \mu^c_{30}\,(\rho-\rho_c)^3 \label{Chemicalpot}
\end{equation}
where $\rho-\rho_c$ and $ T-T_c$ denote the difference between the
density $\rho$ and the Kelvin temperature $T$ from their values
$\rho_c,\, T_c$ at the critical point. This simple  expansion
depends on
 coefficients
corresponding to the values of the partial derivatives of $\mu$  at
the critical point; that is to say,
\begin{equation*}
    \mu^c_{ij} = \frac{\partial^{i+j}
    \mu}{\partial\rho^i\partial T^j} (\rho_c,T_c)
\end{equation*}
The chemical potential $\mu$ has been expanded using a cubic
polynomial in terms of density and temperature; they are  not the
natural variables but  the most convenient for the calculations.\\
 Denoting the coefficients by $A \equiv\mu^c_{11}>0$,
$\displaystyle B \equiv\frac{1}{6}\,\mu^c_{30}>0$ (the sign
conditions are  at the critical point) and $\mu_o \equiv\mu^c +
\mu^c_{01}\, (T-T_c)$ depending only on $T$, Eq. (\ref{Chemicalpot})
reads
\begin{equation}
\mu = \mu_o - A(T_c-T)(\rho-\rho_c)+B(\rho-\rho_c)^3
\label{chemicalpot2}
\end{equation}
Expression (\ref{chemicalpot2}) allows us to solve the problems of
waves in an universal way where $T-T_c$ plays the role of an order
parameter. Such a chemical potential expansion agrees with the van
der Waals equation of state \cite{vanderWaals} where only the values
$A$ and $B$ are
representative of a particular fluid.\\
Consequently, the free energy per unit volume $\psi$ can be written
in the form
\begin{equation}
    \psi=\mu_o\rho-\frac{A}{2}(T_c-T)(\rho-\rho_c)^2+
    \frac{B}{4}(\rho-\rho_c)^4-f(T)
\label{freeenergy}
\end{equation}
where $f$ is an additive function of $T$ only. The pressure $p$
verifies $\ p= \rho\mu-\psi\ $ and we get near $\rho_c$ the
expansion
\begin{equation}
    p = p_c + D(T-T_c) -
    (v-v_c)\Big(A_1(T-T_c)+B_1(v-v_c) \Big)\label{pressure1}
\end{equation}
where $\displaystyle v=\frac{1}{\rho}\ $ denotes the specific
volume, $\displaystyle D = \frac{\partial p}{\partial
T}\,(v_c,T_c),\ f(T_c) = p_c$ and $A_1 =A\,\rho_c^3, \ B_1 =
B\,\rho_c^7$.\\
Due to the general properties of two-phase regions near the critical
point the two-density version of specific internal energy of the
fluid is expressible in the   form \cite{Rowlinson}
\begin{equation}
  \alpha(v,s) = (x^2-y)^2+y^2+\beta\, x+\gamma\, y +\delta \label{Internal}
\end{equation}
\begin{equation*}
 {\rm with} \ \ \  x = a\, r + b\, \sigma, \ \ y = c\, r + d\, \sigma\ \ \
    {\rm where}\ \ r = v-v_c\, , \ \sigma=s-s_c\, ,
\end{equation*}
$\beta, \gamma, \delta, a, b, c, d$ are constants and $s$
denotes the specific entropy of the fluid. \\
Taking into account that $\displaystyle p=-\,\frac{\partial \alpha}{
\partial v}\, (v,s)$,  $ \displaystyle T= \frac{\partial \alpha}{
\partial s}\, (v,s) $ and denoting $\displaystyle\alpha_1(v,s) \equiv \alpha
(v,s)+p_c(v-v_c) - T_c(s-s_c) - \frac{B_1}{4}\, r^4$, we obtain the
partial differential equation
\begin{equation*}
    \frac{\partial \alpha_1}{\partial r} +
    (D-A_1 r) \frac{\partial \alpha_1}{\partial \sigma} =0
\end{equation*}
which yields the general solution for the specific internal energy
\begin{equation*}
    \alpha = g(\sigma-D r + \frac{A_1}{2}\, r^2) + B_1 \frac{r^4}{4}-p_c\, r
     + T_c\, \sigma
\end{equation*}
Due to the fact that $\alpha$ must verify the general form
(\ref{Internal}),   $g(u)$ is necessarily a polynomial with respect
to
 $\displaystyle u = \sigma-D r + \frac{A_1}{2}\, r^2$ in the form $g(u) = E u + G u^2$. The term
 $G u^2$ yields a term of four degree  in $r$ and $r$ must be a function of $x$ only. No term in
 the form $x y$ appears in (\ref{Internal}); consequently $\sigma- D r$ does not contain the
 variable $y$ and the general form of $\alpha$ is
 \begin{equation*}
    \alpha = E(k_2\,y +\frac{A_1}{2}\,k_1^2\,x^2)+G(k_2\,y+\frac{A_1}{2}\,k_1^2\,x^2)^2+
    \frac{B_1}{4}\,
    k_1^4\,x^4-p_c\, r +T_c\, \sigma
 \end{equation*}
where $E$ and $G$ are two constants.\\
 By straightforward
calculations, an identification with the expression
 (\ref{Internal}) yields
 the general expression for the specific internal energy
 \begin{equation}
 \alpha(v,s) = x^4-2x^2y+2y^2-p_c(v-v_c)+T_c(s-s_c)
 \label{specificenergy}
 \end{equation}
 \begin{equation*}
   {\rm with}\ \ x =\sqrt[4]{\frac{B_1}{2}}\ (v-v_c) \ \ {\rm and}\ \ y =\frac{1}{A_1}
   \sqrt{\frac{B_1}{2}}\ [D(v-v_c)-(s-s_c)]
 \end{equation*}
As a consequence the pressure gets the form
 \begin{equation}
p = p_c + 4 \sqrt[4]{\frac{B_1}{2}}\ (y-x^2)\, x +
\frac{D}{A_1}\sqrt{2B_1}\ (x^2-2y)
 \label{pressure2}
 \end{equation}
 and in the case where $s$ is constant, the specific enthalpy reduces to the
 form
 \begin{equation}
h = \frac{DB_1}{A_1}\,
v_c^3\left(v_c^2(\rho-\rho_c)^2+\frac{2D}{A_1}\,(\rho-\rho_c)\right)+
h_o
 \label{enthalpy}
 \end{equation}
 where $h_o$ is a constant depending on $s$.\\
Near the critical point, a fluid behaves like a gas but with a high
density \cite{Rocard}. As for interfaces separating two bulks, the
fluid is not homogeneous. The view that a non-homogeneous fluid near
its critical point may be treated as matter in bulk with a local
energy density that is that of a hypothetically uniform fluid of
composition equal to the local composition with an additional term
arising from the non-uniformity, and that the latter may be
approximated by a gradient expansion typically truncated in second
order is most likely to be successful and perhaps even qualitatively
accurate \cite{Rowlinson,Malomed}. The first study has been done on
the theory of the near-critical interface within the framework of
the van der Waals theory of capillarity \cite{Widom};  the simplest
model able to take into account the density and its gradient uses a
unique supplementary quantity represented by the constant $C$ of
internal capillarity. In S.I. units, the value of $C$ for water at
$20^o$ Celsius   is of the order of $10^{-16}$ (see
\cite{Casal1988}). In the mean field approximation the expression of
the internal energy $\alpha_{nh}$ of a non homogeneous fluid near
the critical point is in the
 form\cite{Rowlinson,Cahn,Casal 1985a}
\begin{equation*}
\alpha_{nh} = \alpha + \frac{C}{2}\,(\nabla \rho)^2
\end{equation*}
The
supplementary term due to the non-homogeneity of the medium in the expression of the internal energy is effective only in interfaces and for fluids
near the critical point.

\section{Motions of a fluid near the critical point}

Equations of motions are classically given in the literature with
the additive {\it second gradient term} \cite{Truskinovsky,Gouin93}.
Fluid motions are associated in the literature with the
\emph{capillary fluid equation of motion} \cite{Casal 1985a}:
\begin{equation}
\rho\, \mathbf{\Gamma} + \nabla p + \rho\,\nabla
(\Omega-C\nabla^2\rho)-\texttt{div}\,  \sigma_v = 0 \label{motion}
\end{equation}
where $p\,$ is the previous thermodynamic pressure associated with
the  medium considered as homogeneous, (i.e. the pressure entering
in the thermodynamic potentials  of the homogeneous fluid near the
critical point), $\Omega$ is the body force potential, $\sigma_v =
\lambda\, \texttt{tr}(\Delta)\, Id + 2\mu\, \Delta $ is the viscous
stress tensor in the classical form where $\Delta$  is the velocity
deformation tensor, $\mathbf{\Gamma}$ the acceleration vector,
$\nabla$ the gradient operator and $\nabla^2$ the Laplacian
operator. The  mass balance yields
\begin{equation}
\frac{\partial \rho}{\partial t} + \texttt{div} \,\rho\, {\mathbf V}
= 0 \label{mass}
\end{equation}
where $ {\mathbf V}$ denotes the velocity of the fluid.\\ Let us
notice that we can obtain the equation of energy such that the
system is compatible with the second law of thermodynamics
\cite{Casal 1985b,Dunn}. Capillary fluids belong to the class of
dispersive systems because the internal energy depends not only on
density but also on its derivatives with respect to space variables;
we have previously seen
 that the constant solutions are nevertheless stable \cite{Gavrilyuk}.\\
Let us study the one-dimensional problem when  the velocity $V$ and
the density $\rho$ are only functions  of the variable $\zeta = z -
c\,t$, where $z$ is the space variable, $t$ the time and $c$ the
wave celerity with respect to a Galilean frame:
\begin{equation*}
    V = V(z - c\,t), \ \  \rho =\rho(z - c\,t)
\end{equation*}
The mass balance equation yields
\begin{equation*}
    -c\, \frac{d\rho}{d\zeta} +\frac{d(\rho\, V)}{d\zeta} = 0
\end{equation*}
and by integrating, we obtain
\begin{equation}
    \rho\,(V-c) = q \label{massbalance}
\end{equation}
where $q$ is constant in the motion. In the cases of waves we obtain
\begin{equation*}
 {\Gamma}  = \frac{1}{2}\,\frac{d}{d\zeta}(V-c)^2
\end{equation*}
We consider that body forces are negligible as in space with
weightlessness. In the uni-dimensional case,
\begin{equation*}
   \texttt{div}\, \sigma_v = (\lambda+2 \, \mu)\, q\, \frac{d^2}{d\zeta^2}\left(\frac{1}{\rho}\right)
\end{equation*}
In the following, we assume that $\nu = (\lambda+2 \, \mu)/\rho$ is
constant in the fluid as assumed in interfaces\cite{Rocard}. Then
Eq. (\ref{motion}) yields
\begin{equation*}
   \frac{d}{d\zeta}\left( \frac{1}{2}\,\frac{q^2}{\rho^2}- C \frac{d^2\rho}{d\zeta^2}-
   \nu\,q\frac{d}{d\zeta}\left(\frac{1}{\rho}\right)+\frac{1}{\rho}\,
   \frac{dp}{d\zeta}\right)=0
\end{equation*}
We consider two cases \emph{a)} isothermal motions corresponding to
fluid motions near equilibrium conditions and \emph{b)} isentropic
motions corresponding to fast velocities. In the two cases, we can
write $\displaystyle \frac{1}{\rho} \frac{dp}{d\zeta} =
\frac{dH}{d\zeta}$ where $H$ is the chemical potential of the fluid
in case \emph{a)} and the specific enthalpy in case \emph{b)}. In
all the cases, near the critical point $\rho\approx\rho_c$;
consequently $\nu\,q/\rho^2\approx \nu\,q/\rho_c^2$ and the equation
of motion becomes
\begin{equation}
C \frac{d^2\rho}{d\zeta^2}-
\frac{\nu\,q}{\rho_c^2}\,\frac{d\rho}{d\zeta}=
H-H_o+\frac{1}{2}\,\frac{q^2}{\rho^2} \label{intequation}
\end{equation}
where $H_o$ is constant.

\section{Example of waves of a fluid near the critical point}
Two main cases are generally considered for traveling waves :
isothermal processes and isentropic processes. Motions can be
conservative as a mathematical limit  of the dissipative case.
Nevertheless, in thin interfaces the viscosity may be neglected and
conservative cases may be considered as   realistic physically
\cite{Langevin}. The cases of viscous fluid involve inequality due
to Liapounov functions.
\subsection{Existence of solitary waves}
Let us notice that for viscous capillary fluid, solitary waves as
kinks cannot appear. Multiplying Eq. (\ref{intequation}) by
$\displaystyle \frac{d\rho}{d\zeta}$  we obtain  by integration :
\begin{equation}
    \frac{C}{2}\,\left(\frac{d\rho}{d\zeta}\right)^2 - K(\rho) + \frac{1}{2}\,\frac{q^2}{\rho}
    =\int_{\zeta_o}^\zeta\frac{\nu\, q}{\rho_c^2}\,
    \left(\frac{d\rho}{d\zeta}\right)^2 d\zeta
    \label{Liapounov}
\end{equation}
where $\zeta_o$ is a constant and $K'(\rho)= H(\rho)-H_o$.\\
Eq. (\ref{Liapounov}) allows to obtain a first integral only when
$\nu=0$ . Due to the fact that
\begin{equation*}
{\rm for}\ \ \nu>0\,\ \  {\rm and}  \ \ \zeta>\zeta_o\,,\ \
\int_{\zeta_o}^\zeta\frac{\nu\,
q}{\rho_c^2}\,\left(\frac{d\rho}{d\zeta}\right)^2 d\zeta >
  0
\end{equation*}
it is not possible to obtain $\rho(-\infty) = \rho(+\infty)$ and no
solitary wave can appear in dissipative motions.
\subsection{Isothermal waves}
In the isothermal case  the wave motion is obtained by using the
chemical potential $\mu$ given by Eq. (\ref{chemicalpot2})
\begin{equation}
\frac{1}{2}\,\frac{d^2\rho}{d\zeta^2}-\frac{\nu\,q}{2\,\rho^2_cC}\,\frac{d\rho}{d\zeta}=
- \frac{A}{2\,C}\,(T_c-T)\,(\rho-\rho_c)+
\frac{B}{2\,C}\,(\rho-\rho_c)^3 + \frac{q^2}{4\,C\,\rho^2}\,+\, k_o
\label{isothermalmotion}
\end{equation}
where $k_o$ is a constant. \\
Near the critical point $|(\rho-\rho_c)/\rho_c | \ll 1$;
 consequently, we get the following expansion to the second order in
 $(\rho-\rho_c)/\rho_c$
\begin{equation*}
    \frac{1}{\rho^2}\approx \frac{1}{\rho_c^2}\left(1-\frac{2(\rho-\rho_c)}{\rho_c}
    +\frac{3(\rho-\rho_c)^2}{\rho_c^2}\right)
\end{equation*}
and to the fourth order Eq. (\ref{isothermalmotion}) yields
\begin{eqnarray*}
   \frac{1}{2}\,\frac{d^2\rho}{d\zeta^2}-\frac{\nu\,q}{2\,\rho_c^2C}\,
   \frac{d\rho}{d\zeta}&=&  \frac{B}{2\,C}\left(\rho-\rho_c+\frac{q^2}{2
    B \rho_c^4}\right)^3\\
   &&-\frac{1}{2\, C}
\left(A(T_c-T)+\frac{q^2}{\rho_c^3}\right)+
\left(\rho-\rho_c+\frac{q^2}{2 B\rho_c^4}\right)+k_1
\end{eqnarray*}
where $k_1$ is constant. We define the following change of
variables:
\begin{equation}
\rho =\left(\rho_c-\frac{q^2}{2\, B\,\rho_c^4}\right)(1+\varepsilon
Y), \  \zeta=L\ \xi \  {\rm and} \ \ Q = \frac{q
\left(A(T_c-T)+\displaystyle
 \frac{q^2}{\rho_c^3}\right)}{(2C)^{3/2} {B}^{1/2}\rho_c^2}
\label{changevariables}
\end{equation}
\begin{equation*}
    {\rm with}\ \ L^2 =\frac{2\, C}{A(T_c-T)+
    \displaystyle\frac{q^2}{\rho_c^3}} \ \ \
    {\rm and}     \ \  \
    \varepsilon^2 =\frac{{A \left(T_c-T\right)+\displaystyle\frac{q^2}{\rho_c^3}}}
{B \left(\rho_c-\displaystyle\frac{q^2}{2 B\rho_c^4}\right)^2}
\end{equation*}
Then, the equation of isothermal waves writes
\begin{equation}
  \frac{1}{2}\,\frac{d^2Y}{d\xi^2} - \nu\, Q \,\frac{dY}{d\xi} = Y^3-Y + k_1 \label{approxisoth}
\end{equation}

\emph{{\bf Conservative case}}, $\nu=0$: we obtain the first
integral
\begin{equation}
\left(\frac{dY}{d\xi}\right)^2 = (1-Y^2)^2 -a_1Y-b_1 \label{curves}
\end{equation}
where $a_1$ and $b_1$ are two constants.  The intersections of the straight line $a_1Y+b_1$ and of the
quartic $(1-Y^2)^2$ yield the density range

\begin{figure}[ht]
\centerline{\epsfxsize=4.1in\epsfbox{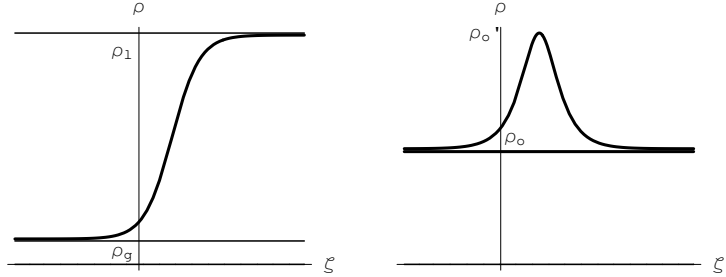}} \caption{On the left
side of the figure a phase transition wave  is represented
(densities $\rho_l$ and $\rho_g$ are associated with the liquid and
vapor bulks); on the right side a traveling wave is a kink
associated with case \emph{a)} (density $\rho_o$ corresponds to the
bulk of the fluid). \label{waves}}
\end{figure}

\emph{Interfacial propagation} - We recall the main results of
(\cite{Gouin93}): in the bulks phases, the densities are constant
and thus the first and second derivatives of $Y$ are zero. The
straight line is tangent to the quartic at the associated points. If
the bulk phases are different on both sides of the interface, the
straight line has to be bitangential to the quartic which implies
$a_1$ and $b_1$ are null (case of two phases). A liquid-vapor
interface wave is similar to the one obtained in the equilibrium
case ($q=0$) but mass flows through the interface. Vaporization or
condensation phenomena depend on the sign of $q$. This case
corresponds to a shock wave in the sense of Slemrod \cite{Slemrod}
and
\begin{equation}
    \rho = \left(\rho_c-\frac{q^2}{2\, B\,\rho_c^4}\right)\left(1+\varepsilon\, {\rm tanh}\left(\frac{\zeta}
    {L}\right)\right) \label{phasetransition}
\end{equation}
(see left side of fig. \ref{waves}).

\emph{Solitary waves} - It is also possible to obtain a solitary
wave moving in one of the bulk phases (liquid or vapor). The
straight line is tangent to the quartic at the point associated with
the bulk $Y=Y_o$ and intersect the quartic at the point associated
with the \emph{middle} of the
 wave $Y=Y'_o$. When $Y$ belongs to the interval between $Y_o$ and $Y'_o$, the right hand side
 of Eq. (\ref{curves}) must be positive and this authorizes only two possibilities:\\
  {\emph a)}   $-1<Y_o<-1/\sqrt{3}\,$ and $\, Y_o<Y'_o<1 $, the density increases in the wave (see right side of fig. \ref{waves}),\\
{\emph b)}  $1/\sqrt{3}<Y_o<1\,$ and $\,-1< Y'_o<Y_o $, the density
decreases in the wave.

\emph{Consequences} - The change of variables
(\ref{changevariables}) assumes that $\displaystyle T_c-T
+\frac{q^2}{A \rho_c^3}$ is positive; Then, for $T>T_c$ the flow
through the interface must be more important than the limit
$\displaystyle q_m =\sqrt{  A \rho_c^3(T-T_c)}$ and the wave
velocity $\ell$ must be greater than the limiting value
$\displaystyle \ell_m =\sqrt{  A \rho_c (T-T_c) }$ which is the
celerity of isothermal
acoustic waves.\\

 \emph{{\bf Dissipative case}}, $\nu>0$ : In this case no
solitary wave appears (see section 4.1). Let us define
$\displaystyle Z(Y) = \frac{dY}{d\xi}$; then Eq. (\ref{approxisoth})
yields
\begin{equation*}
    \frac{1}{2}\, Z Z' -\nu\, Q\, Z = Y^3 -Y + k_1
\end{equation*}
We look for the phase transition waves with a solution in the form
\begin{equation*}
    Z(Y) = \alpha_1 Y^2 + \beta_1 Y + \gamma_1
\end{equation*}
where the polynomial $\alpha_1 Y^2 + \beta_1 Y + \gamma_1$  has two real roots. Then, the
 differential equation
 \begin{equation*}
   \frac{dY}{d\xi} = \alpha_1 Y^2 + \beta_1 Y + \gamma_1
\end{equation*}
has a solution in the same form than (\ref{phasetransition}); (see
left side of fig. \ref{waves}).\\ Straightforward calculations prove
that the polynomial has two real roots when
\begin{equation*}
   \nu^2 q^2 \left(A \left(T_c-T\right)+\frac{q^2}{\rho_c^3}\right) \leq 24\,  C^3 B \rho_c^4
\end{equation*}
which corresponds to a velocity $q/\rho_c$ of the wave small enough.
This condition  does not appear in the conservative case where
$\nu=0$.
\subsection{Isentropic waves}
\begin{figure}[ht]
\centerline{\epsfxsize=5.1in\epsfbox{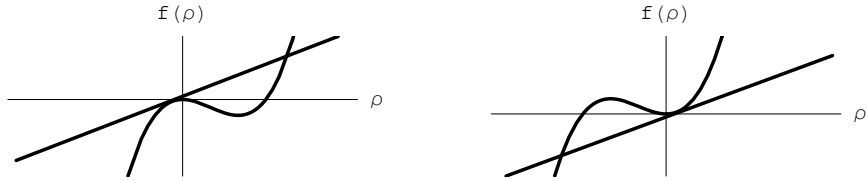}} \caption{On the left
side of the figure, the case where the line $f(Y) = a_o Y + b_o$ is
above the cubic representing the left side of Eq.
(\ref{isentropicmotion}) is  presented; on the right side  the
opposite situation: the line $f(Y) = a_o Y + b_o$ is below the
cubic. \label{cubic}}
\end{figure}

 For an isentropic conservative motion, $s=Cte$ and the equation of
motion is in a form deduced from Eq. (\ref{intequation}),
\begin{equation*}
    C \frac{d^2\rho}{d\zeta^2}= h-h_o +\frac{1}{2}\,\frac{q^2}{\rho^2}
\end{equation*}
where $h$ is now the specific enthalpy. We consider the new change
of variables
\begin{equation*}
\rho = \rho_c\,(1+Y),\ \ \xi = L\,\zeta \ \ \ {\rm and}\ \ \ q =
b\,Q
\end{equation*}
with
\begin{equation*}
    L^2 = \frac{2\, C \rho_c^4\, A_1}{D\, B_1}, \ \ \
    b^2 = \frac{D\, B_1}{\rho_c\, A_1}
\end{equation*}
and the equation of wave motions yields
\begin{equation}
    \frac{1}{4}\,\left(\frac{dY}{d\xi}\right)^2 =
    \frac{Y^3}{3}\,(1+3\,Q^2) + \frac{Y^2}{2}\,(\tau-2\,Q^2)+ a_oY
    + b_o \label{isentropicmotion}
\end{equation}
with $\displaystyle \tau = \frac{2\,\rho_c\,D}{A_1}$ and $a_o$ and
$b_o$ are two constants. We denote by $\displaystyle \ell_m =
\frac{2\,D^2\,B_1}{A_1^2}\,v_c^2\ $
 a limit celerity of the waves with respect to the fluid.
 Due to the form of the second member of Eq. (\ref{isentropicmotion}), we notice
 immediately that it is not possible to obtain phase transition
 waves.
 Consequently
 traveling waves cannot appear in dissipative motions when
 the entropy is constant in the fluid. The two
cases are represented on fig  \ref{cubic}.\\

 \emph{Case 1}: $Q^2< \tau/2\ $ or $\ \ell < \ell_m$. In this case solitary
 waves cannot appear (see left side of fig  \ref{cubic}).

 \emph{Case 2}: $Q^2> \tau/2\ $ or $\ \ell > \ell_m$. In this case solitary
 waves are possible depending on initial conditions (see right side of fig  \ref{cubic}).
\\
In the van der Waals model of pressure we obtain by straightforward
calculations
\begin{equation*}
    D = \frac{4\, p_c}{T_c}, \ \ \ A_1 = \frac{6\, p_c \rho_c}{T_c},
    \ \ \ B_1 = \frac{3\, p_c\,\rho_c^3}{2}
\end{equation*}
and consequently, $\displaystyle \ell_m =
\sqrt{\frac{4}{3}\,\frac{p_c}{\rho_c}}$\ .

\subsection{Fluid velocity near the critical point} For a given
density, the fluid velocity is deduced at time $t=0$ from Eq.
({\ref{massbalance}). Consequently,
\begin{equation*}
    V(z) = c+\frac{q}{\rho(z)}
\end{equation*}
where $V(z)$ is the fluid velocity at $z$ and $c$ is an arbitrary
constant. At any time t,
\begin{equation*}
    V(z-c\,t) = c+\frac{q}{\rho(z-c\,t)}
\end{equation*}
where initial conditions yield the arbitrary velocity $c$.

\emph{Interface propagation} - In a phase transition wave, the fluid
changes from liquid to vapor as its volume increases. Consequently
such a phenomenon cannot occur in a closed tube but only in a tube
with only one closed end. For example, the tube is closed at the
other end with a piston whose displacement is imposed. The fluid
velocity at the fixed end is zero (for example liquid bulk) and  $c
= -q/\rho_l $. If we impose a value $U$ for the piston velocity in
the vapor bulk, we deduce the value of $q$
\begin{equation*}
 q\left(\frac{1}{\rho_v}-\frac{1}{\rho_l}\right) = U
\end{equation*}
where $\rho_v$ and $\rho_l$ are the values of the density in the
vapor and liquid bulks. From the values of $\rho_v$ and $\rho_l$
deduced from expression (\ref{phasetransition}), we determine the
flow $q$ and the velocities $\ell\simeq q/\rho_c$ and $c$.

\emph{Solitary wave} - The volume of the interface that moves in
only one bulk phase is constant. Such a wave may be moving in a
closed tube such as Natterer tube \cite{Bruhat} . At the ends
(assumed far from the region of the wave) the velocity in the bulk
phase is zero and $c\simeq -q/\rho_c$.  When the temperature is
close to $T_c$, the model of fluid endowed with internal capillarity
allows to obtain traveling waves in a tube which depends on two
arbitrary parameters.

\section{Conclusion}
The mean field approximation for fluid near the critical point is
able to predict solitons and transition of phase waves. Two kinds of
waves are investigated: \emph{a)} liquid-vapor waves in the case of
isothermal medium whose the celerity depends on the distance between
the temperature and its critical value. Above the critical
temperature, the fluid behaves like an elastic medium and the waves
are supersonic with respect to the isothermal sound velocity
\cite{Gouin93}. Below the critical temperature, the fluid behaves
like a nonrigid medium and the wave velocities can take any value in
conservative motion but are bounded in dissipative motion
proportionally to the viscosity of the fluid. \emph{b)} Traveling
kinks appear in the
 conservative case as well as for isothermal than for
isentropic motions.\\
Thermodynamics functions and parameters vanish or diverge at the
critical point proportionally to some power of the distance from
that point currently measured as $T-T_c$. The critical exponents are
central to the discussion of critical phenomena and can be
generalized by nonclassical value of critical point exponents for
the potentials given in section 2.  The problem can be extended for
multi-component fluid mixtures but  critical points are not unique;
for a mixture of fluids there is a curve of critical points.

\section*{Acknowledgments}The paper has been partially supported by
PRIN 2000 (Coordinator Prof. T. Ruggeri) and by G.d.R. CNES/CNRS
2258.

\end{document}